\documentclass[12pt]{article}
\usepackage{amssymb}
\usepackage{amsmath}
\usepackage{slashed}

\newcommand{\G}[1]{ \ensuremath{ \Gamma_{#1} } }
\newcommand{\GUp}[1]{ \ensuremath{ \Gamma^{#1} } }
\newcommand{\ie}{{\it i.e.}}

\newcommand{\e}{\begin{equation}}
\newcommand{\ee}{\end{equation}}
\newcommand{\ea}{\begin{eqnarray}}
\newcommand{\eea}{\end{eqnarray}}

\begin{document}

\begin{flushright}
\end{flushright}
\begin{flushright}
%{\sf \today}
\end{flushright}
\begin{center}

{\LARGE {\sc M2-Branes and Background Fields\\}}

\bigskip

{\sc Neil Lambert\footnote{neil.lambert@kcl.ac.uk} and Paul Richmond\footnote{paul.richmond@kcl.ac.uk}} \\
{Department of Mathematics\\
King's College London\\
The Strand\\
London WC2R 2LS, UK\\}

\end{center}

\bigskip
\begin{center}
{\bf {\sc Abstract}}
\end{center}

We discuss the coupling of multiple M2-branes to the background 3-form and 6-form gauge fields of
eleven-dimensional supergravity, including the coupling of the Fermions. In particular we show in detail how a
natural generalization of the Myers flux-terms, along with the resulting curvature of the background metric,
leads to mass terms in the effective field theory.

\newpage

\section{\sl Introduction}

For a single M2-brane propagating in an eleven-dimensional spacetime with
coordinates $x^m$ the full non-linear effective action including
Fermions and $\kappa$-symmetry was obtained in
\cite{Bergshoeff:1987cm}. The Bosonic part of the effective action
is
\begin{eqnarray}\label{BSTaction}
\nonumber S &=& -T_{M2}\int d^3\sigma  \sqrt{-{\rm det}(\partial_\mu
x^m\partial_\nu x^n g_{mn})}\\&&\hskip2cm  + \frac{T_{M2}}{3!}\int
d^3\sigma \, \epsilon^{\mu\nu\lambda}\partial_\mu x^m \partial_\nu
x^n\partial_\lambda x^p C_{mnp}\ .
\end{eqnarray}
Here $C_{mnp}$ is the M-theory 3-form potential,  $g_{mn}$ the
eleven-dimensional metric and $T_{M2}\propto M_{pl}^3$ is the
M2-brane tension.

If we go to static gauge, $\sigma^\mu = x^\mu$, $\mu=0,1,2$ then the
M2-brane has world-volume coordinates $x^\mu$  and the $x^I$,
$I=3,4,5,....,10$ become 8 scalar fields. In this paper we will be
interested in the lowest order terms in an expansion in the
eleven-dimensional Planck scale $M_{pl}$. In this case the
canonically normalized scalars are $X^I = x^I/\sqrt{T_{M2}}$. These
have mass-dimension $1/2$ whereas $g_{mn}$ and $C_{mnp}$ are
dimensionless.

We next seek a generalization of this action to lowest order in $M_{pl}$ but for multiple M2-branes. The
generalization of the first term in (\ref{BSTaction}) was first proposed in
\cite{Bagger:2006sk}\cite{Gustavsson:2007vu}\cite{Bagger:2007jr}\cite{Bagger:2007vi}. This has the maximal
${\cal N}=8$ supersymmetry and describes two M2-branes in an ${\mathbb R}^8/{\mathbb Z}_2$ orbifold
\cite{Lambert:2008et}\cite{Distler:2008mk} but cannot be extended to more M2-branes
\cite{Gauntlett:2008uf}\cite{Papadopoulos:2008sk} (although there are interesting models with Lorentzian
signature on the 3-algebra \cite{Gomis:2008uv}\cite{Benvenuti:2008bt}). It was then further generalized in
\cite{Aharony:2008ug}\cite{Aharony:2008gk} for arbitrary M2-branes and manifest ${\cal N}=6$ supersymmetry in
an ${\mathbb R}^8/{\mathbb Z}_k$ orbifold.

In this paper we will obtain the generalization of the second term (\ie\ the Wess-Zumino term) which gives the
coupling of the M2-branes to background gauge fields. In the well studied case of D-branes, where the low
energy effective theory is a maximally supersymmetric Yang-Mills gauge theory with fields in the adjoint
representation, the appropriate generalization was given by Myers \cite{Myers:1999ps}. In the case of multiple
M2-branes  the scalar fields $X^I$ and Fermions now take values in a 3-algebra which carries a bifundamental
representation of the gauge group. Thus we wish to adapt the Myers construction to M2-branes. For alternative
discussions of the coupling of multiple M2-branes to background fields see \cite{Li:2008eza},\cite{Kim:2009nc}.

The rest of this paper is as follows. In section 2 we will discuss
the relevant couplings, to lowest order in $M_{pl}$, for the ${\cal
N}=8$ Lagrangian of \cite{Bagger:2007jr} and demonstrate that, by an
appropriate choice of terms, the action is local and gauge
invariant. We will also supersymmetrize the case where the
background field $G_{IJKL}$ is non-vanishing and demonstrate that
this leads to the mass-deformed theories first proposed in \cite{Hosomichi:2008qk}\cite{Gomis:2008cv}.
In section 3 we will repeat our analysis for the case of ${\cal
N}=6$ supersymmetry, which includes both the ABJM
\cite{Aharony:2008ug} and ABJ \cite{Aharony:2008gk} models leading to the mass deformed models of \cite{Hosomichi:2008jb},\cite{Gomis:2008vc}. In
section 4 we will discuss the physical origin of the flux-squared
term that arises by supersymmetry. In particular we will demonstrate
that this term arises via back reaction of the fluxes which leads to
a curvature of spacetime. Section 5 will conclude with a
discussion of our results.

\section{\sl ${\cal N}=8$ Theories}

Let us first consider the maximally supersymmetric case. We follow
the notion and conventions of \cite{Bagger:2008se}. Although this
case has only been concretely identified with the effective action
of two M2-branes in an ${\mathbb R}^8/{\mathbb Z}_2$ orbifold
\cite{Lambert:2008et}\cite{Distler:2008mk} it is simpler to handle
and hence the presentation is clearer. In the next section we will
repeat our analysis for the case of ${\cal N}=6$.

\subsection{\sl Non-Abelian Couplings to Background Fluxes}

 The scalars
$X^I$ live in a 3-algebra with totally anti-symmetric triple product $[X^I,X^J,X^K]$ and invariant inner
product ${\rm Tr}(X^I,X^J)$ subject to a quadratic fundamental identity and the condition that ${\rm
Tr}(X^I,[X^J,X^K,X^L])$ is totally anti-symmetric in $I,J,K,L$ \cite{Bagger:2007jr}. An important distinction
with the usual case of D-branes based on Lie algebras is that Tr is an inner-product and not a map from the
Lie algebra to the real numbers. In particular there is no gauge invariant object such as ${\rm Tr}(X^I)$. Thus
the only gauge-invariant terms that we can construct involve an even number of scalar fields.

In this paper  we wish to consider the decoupling limit
$T_{M2}\to\infty$ since, unlike String Theory, there are no other
parameters that we can tune to turn off the coupling to gravity. In
particular it is not clear to what extent finite $T_{M2}$ effects
can be consistently dealt with in the absence of the full
eleven-dimensional dynamics.

Assuming that there is no metric dependence
we start with the most general form for a non-Abelian pull-back of the background gauge fields
to the M2-brane world-volume:
\begin{eqnarray}
\nonumber  S_{C} &=& \frac{1}{3!}\epsilon^{\mu\nu\lambda}\int
d^3 x \Big( aT_{M2} C_{\mu\nu\lambda}+3b C_{\mu IJ} \, {\rm Tr}
  (D_\nu X^I,D_\lambda X^J)  \\
\nonumber     && + 12cC_{\mu\nu IJKL} \, {\rm Tr}(D_\lambda
   X^I,[X^J,X^K,X^L])\\
   && + 12dC_{[\mu IJ}C_{\nu KL]} \, {\rm Tr}(D_\lambda
   X^I,[X^J,X^K,X^L])+\ldots
 \Big)
\end{eqnarray}
where $a,b,c,d$ are dimensionless constants that we
have included for generality and the ellipsis denotes terms that are
proportional to negative powers of $T_{M2}$ and hence vanish in the
limit $T_{M2}\to \infty$.

Let us make several comments. First note that we have allowed the
possibility of higher powers of the background fields. In D-branes
the Myers terms are linear in the R-R fields however they also
include non-linear couplings to the NS-NS 2-form. Since all these
fields come from the M-theory 3-form or 6-form this suggests that we
allow for a non-linear dependence in the M2-brane action.

Note that gauge invariance has ruled out any terms where the
$C$-fields have an odd number of indices that are transverse to the
M2-branes (although the last term could have a part of the form $C_{\mu\nu I }C_{JKL}$).
This is consistent with the observation that the ${\cal
N}=8$ theory describes M2-branes in an ${\mathbb R}^8/{\mathbb Z}_2$
orbifold and hence we must set to zero any components of $C_3$ or
$C_6$ with an odd number of $I,J$ indices.

The first term is the ordinary coupling of an M2-brane to the
background 3-form and hence we should take $a=N$ for $N$ M2's. The
second line leads to a non-Lorentz invariant modification of the
effective 3-dimensional kinetic terms. It is also present in the
case of a single M2-brane action (\ref{BSTaction}) where we find
$b=1$ which we will assume to be the case in the non-Abelian theory.\footnote{This is an assumption since  the overall centre of mass
zero mode $x^\mu$ that appears in (\ref{BSTaction}) is absent in the
non-Abelian generalizations.} The final
term proportional to $d$ in fact vanishes as ${\rm Tr}(D_\lambda
   X^{[I},[X^J,X^K,X^{L]}])= \frac{1}{4}\partial_\lambda{\rm Tr}(X^I,[X^J,X^K,X^L])$ which is symmetric under
   $I,J \leftrightarrow K,L$. Thus we can set
$d=0$.

Finally note that we have allowed the M2-brane to couple to both the
3-form gauge field and  its electromagnetic 6-form dual defined by
$G_4=dC_3$, $G_7=dC_6$ where
\begin{equation}
G_7  = \star G_4
 - \frac{1}{2}C_3\wedge G_4\ .
\end{equation}
The equations of motion of eleven-dimensional supergravity imply
that $dG_7=0$. However $G_7$ is not gauge invariant under $\delta
C_3 = d\Lambda_2$. Thus $S_{C}$ is not obviously gauge invariant or
even local as a functional of the
 eleven-dimensional gauge fields. As such one should integrate by
parts whenever possible and seek  to find an expression which is
manifestly gauge invariant.

To discuss the gauge invariance under $\delta C_3=d\Lambda_2$ we
first integrate by parts and discard all boundary terms
\begin{eqnarray}
\nonumber  S_{C} &=& \frac{1}{3!}\epsilon^{\mu\nu\lambda}\int d^3 x
\Big(
NT_{M2} C_{\mu\nu\lambda}\\
&&+\frac{3}{2} G_{\mu\nu IJ} \, {\rm Tr}
  (X^I,D_\lambda X^J) -\frac{3}{2}C_{\mu IJ} \, {\rm Tr}
  ( X^I,\tilde F_{\nu\lambda}X^J)\\
\nonumber   && - c G_{\mu\nu\lambda IJKL} \, {\rm Tr}(
   X^I,[X^J,X^K,X^L])\Big)\ .
\end{eqnarray}
Here we have used the fact that $C_{\mu\nu I}$ and $C_{\mu\nu\lambda
IJK}$ have been projected out by the orbifold and hence $G_{\mu\nu
IJ} =2
\partial_{[\mu} C_{\nu ]IJ}$ and $G_{\mu\nu\lambda IJKL} = 3\partial_{[\mu}C_{\nu\lambda] IJKL}$.

We find a coupling to the world-volume gauge field strength $\tilde
F_{\nu\lambda}$ but this term is not invariant under the gauge
transformation $\delta C_3 = d\Lambda_2$. However it can be cancelled
by adding the term
\begin{equation}
S_{F}=\frac{1}{4}\epsilon^{\mu\nu\lambda}\int d^3 x \, {\rm Tr}(X^I,\tilde  F_{\mu\nu} X^J)
C_{\lambda
IJ}\ ,
\end{equation}
to $S_{C}$. Such terms involving the world-volume gauge field
strength also arise in the  action of multiple D-branes.

Next consider the terms on the third line. Although $G_7$ is not
gauge invariant $G_7+\frac{1}{2}C_3\wedge G_4$ is. Thus we also add the term
\begin{equation}
S_{CG}=-\frac{c}{2\cdot 3!}\epsilon^{\mu\nu\lambda}\int d^3 x \, {\rm Tr}(X^I,[X^J,X^K,X^L])(C_3\wedge G_4)_{\mu\nu\lambda IJKL}\ .
\end{equation}
and obtain a gauge invariant action.

%Note that one might object that, due to the orbifold projection, we must have $C_{IJK}=0$  and hence
%$G_{IJKL}=4\partial_{[I} C_{JKL]}=0$. However if allow for a dual 6-form potential with $C_{\mu\nu IJKL}\ne 0$
%then $G_{\mu\nu\lambda IJKL}=3\partial_{[\mu} C_{\mu\nu] IJKL}$ and hence $G_{IJKL}$ can be non-vanishing.

To summarize we find that the total flux terms are, in the limit $T_{M2}\to \infty$,
\begin{eqnarray}
\nonumber   S_{flux} &=& S_C+S_F+S_{CG} \\
   &=& \frac{1}{3!}\epsilon^{\mu\nu\lambda}\int d^3
x \Big( NT_{M2} C_{\mu\nu\lambda}+\frac{3}{2} G_{\mu\nu IJ} \, {\rm Tr}
  (X^I,D_\lambda X^J)\\
\nonumber   && - c(G_7+\frac{1}{2}C_3\wedge G_4)_{\mu\nu\lambda
IJKL} \, {\rm Tr}(
   X^I,[X^J,X^K,X^L])\Big)\ .
\end{eqnarray}
In section 4 we will argue that $c=2$.

\subsection{\sl Supersymmetry}

In this section we wish to supersymmetrize the flux term $S_{flux}$
that we found above. There are also similar calculations in \cite{Grana:2002tu}\cite{Marolf:2003vf}\cite{Camara:2003ku}
where the flux-induced Fermion masses on D-branes were obtained. Here we will be
interested in the final term since only it preserves
3-dimensional Lorentz invariance (the first term is just a constant
if it is Lorentz invariant). Thus for the rest of this section we will
consider backgrounds where
\begin{equation}\label{Sflux}
\mathcal{L}_{flux} = c\tilde G_{IJKL} \, {\rm Tr}(X^I,[X^J,X^K,X^L])\ ,
\end{equation}
with
\begin{eqnarray}
\nonumber \tilde G_{IJKL} &=&-
\frac{1}{3!}\epsilon^{\mu\nu\lambda}(G_7+\frac{1}{2}C_3\wedge G_{4})_{\mu\nu\lambda
IJKL}\\ &=&
\frac{1}{4!}\epsilon_{IJKLMNPQ}G^{MNPQ}
\end{eqnarray}
and $G_{IJKL}$ is assumed to be constant.

To proceed we take the ansatz for the Lagrangian in the presence of background fields
to be
    \begin{equation}
        \mathcal{L} = \mathcal{L}_{\mathcal{N}=8} + \mathcal{L}_{mass} + \mathcal{L}_{flux}\ ,
    \end{equation}
where $\mathcal{L}_{\mathcal{N}=8}$ is the Lagrangian detailed in
\cite{Bagger:2007jr},
    \begin{align}
        \mathcal{L}_{mass} &= -\frac{1}{2}m^2\delta_{IJ} \, {\rm Tr} (X^{I} , X^{J} ) + b \, {\rm Tr} ( \bar{\Psi} \GUp{IJKL} , \Psi ) \tilde{G}_{IJKL}\ ,
    \end{align}
and $m^2$ and $B$ are constants. We use conventions where $\Psi$ and $
\epsilon$ are eleven-dimensional spinors satisfying the constraints
$\G{012} \Psi = - \Psi$ and $\G{012} \epsilon = \epsilon$.

As shown in \cite{Bagger:2007jr}, $\mathcal{L}_{\mathcal{N}=8}$ is invariant
under the supersymmetry transformations
    \begin{align}
        \delta X^{I}_{a} &= i \bar{\epsilon} \GUp{I} \Psi_{a} \nonumber \\
        \delta \tilde{A}_{\mu}{}^{b}{}_{a} &= i \bar{\epsilon} \G{\mu} \G{I} X^{I}_{c} \Psi_{d} f^{cdb}{}_{a}  \\
        \delta \Psi_{a} &= D_{\mu}X^{I}_{a}\GUp{\mu} \GUp{I} \epsilon - \frac{1}{6} X^{I}_{b} X^{J}_{c} X^{K}_{d} f^{bcd}{}_{a} \GUp{IJK} \epsilon \nonumber\ .
    \end{align}
We propose additional supersymmetry transformations of the following
form
    \begin{align}
        \delta ' X^{I}_{a} &= 0 \nonumber \\
        \delta ' \tilde{A}_{\mu}{}^{b}{}_{a} &= 0 \\
        \delta ' \Psi_{a} &= \omega \GUp{IJKL} \GUp{M} \epsilon X^{M}_{a} \tilde{G}_{IJKL} \nonumber\ ,
    \end{align}
where $\omega$ is a real dimensionless parameter.

Applying the supersymmetry transformations to the mass deformed
Lagrangian gives
    \begin{eqnarray}
        \tilde{\delta} \mathcal{L} &=& ( \delta ' + \delta ) ( \mathcal{L}_{\mathcal{N}=8} + \mathcal{L}_{mass} + \mathcal{L}_{flux} ) \\
        &=& ( i\omega + 2b ) \, {\rm Tr} ( \bar{\Psi} \GUp{\mu} \GUp{MNOP} \GUp{I} \epsilon , D_{\mu} X^{I} ) \tilde{G}_{MNOP} \nonumber \\
        && + \frac{i \omega}{2} \, {\rm Tr} ( \bar{\Psi} \GUp{IJ} \GUp{MNOP} \GUp{K} \epsilon , [ X^{I} , X^{J} , X^{K} ] ) \tilde{G}_{MNOP} \nonumber \\
        && - \frac{2b}{6} \, {\rm Tr} ( \bar{\Psi} \GUp{MNOP} \GUp{IJK} \epsilon, [ X^{I} , X^{J} , X^{K} ] ) \tilde{G}_{MNOP} \nonumber \\
        && + 4ic \, {\rm Tr} ( \bar{\Psi} \GUp{I} \epsilon , [ X^{J} , X^{K} , X^{L} ] ) \tilde{G}_{IJKL} \nonumber \\
        && + i m^{2} \delta_{IJ} \, {\rm Tr} ( \bar{\Psi} \GUp{I} \epsilon , X^{J} ) \nonumber \\
        && + 2b\omega \, {\rm Tr} ( \bar{\Psi} \GUp{IJKL} \GUp{MNOP} \GUp{Q} \epsilon , X^{Q} ) \tilde{G}_{IJKL} \tilde{G}_{MNOP}\ .
    \end{eqnarray}
To eliminate the term involving the covariant derivative we must set
$b=\nolinebreak-i\omega/2$. Substituting for $b$, expanding out the
gamma matrices and using anti-symmetry of the indices yields
    \begin{eqnarray}
        \tilde{\delta} \mathcal{L} &=& \frac{2i\omega}{3} \, {\rm Tr} ( \bar{\Psi} \GUp{IJKMNOP} \epsilon , [ X^{I} , X^{J} , X^{K} ] ) \tilde{G}_{MNOP} \nonumber \\
        && + (4ic-16i\omega) \, {\rm Tr} ( \bar{\Psi} \GUp{L} \epsilon , [ X^{I} , X^{J} , X^{K} ] ) \tilde{G}_{LIJK} \nonumber \\
        && + i m^{2} \delta_{IJ} \, {\rm Tr} ( \bar{\Psi} \GUp{I} \epsilon , X^{J} ) \nonumber \\
        && - i \omega^{2} \, {\rm Tr} ( \bar{\Psi} \GUp{JKLM} \GUp{NOPQ} \GUp{I} \epsilon , X^{I} ) \tilde{G}_{JKLM} \tilde{G}_{NOPQ}\ .
    \end{eqnarray}
Defining $\slashed{\tilde G} = \tilde{G}_{JKLM}\GUp{JKLM} $ and using
Hodge duality of the gamma matrices leads to
    \begin{eqnarray}
        \tilde{\delta} \mathcal{L} &=& \frac{96i\omega}{6} \left(-1 + \frac{c}{4\omega} - \star \right) \tilde{G}_{LIJK} \, {\rm Tr} ( \bar{\Psi} \GUp{L} \epsilon , [ X^{I} , X^{J} , X^{K} ] ) \nonumber \\
        && + i \, {\rm Tr} ( \bar{\Psi} \left( m^{2} - \omega^{2} \slashed{\tilde G} \slashed{\tilde G} \right) \GUp{I} \epsilon , X^{I} )\ .
    \end{eqnarray}
Invariance then follows if the following equations hold
    \begin{equation} \label{1stOrderCondition}
        \left(-1 + \frac{c}{4\omega} - \star \right) \tilde{G}_{LIJK} = 0
\qquad{\rm and}\qquad        \left( m^{2} - \omega^{2}
\slashed{\tilde G} \slashed{\tilde G} \right) \GUp{I} \epsilon = 0\ .
    \end{equation}
Since we assume that $c\ne 0$, the first equation implies
$\omega=c/8$ and $\tilde{G}$ is self-dual. It follows from the
result $\GUp{3456789(10)} \slashed{\tilde G} = \slashed{\tilde G}$
that the second equation is satisfied by
    \begin{equation}
        \slashed{\tilde G} \slashed{\tilde G} =  \frac{32m^2}{c^{2}} \left( 1 + \GUp{3456789(10)} \right)\ .
    \end{equation}
Expanding out the left hand side and using the self-duality of
$\tilde{G}$ one sees that this is equivalent to the two conditions
    \begin{equation}\label{Gcon}
        m^2 = \frac{c^2}{32\cdot 4!} G^2  \qquad {\rm and}\qquad  G_{MN[IJ}
        G_{KL]}{}^{MN}=0\ ,
    \end{equation}
where $G^2 = G_{IJKL}G^{IJKL}$.

The superalgebra can be shown to close on-shell. We first consider
the gauge field and find that the transformations close into the
same translation and gauge transformation as in the un-deformed
theory;
    \begin{align}
        [ \tilde{\delta}_{1} , \tilde{\delta}_{2} ] \tilde{A}_{\mu}{}^{b}{}_{a} &= [ \delta_{1} + \delta_{1}' , \delta_{2} + \delta_{2}' ] \tilde{A}_{\mu}{}^{b}{}_{a} \nonumber \\
            &= v^{\nu} \tilde{F}_{\mu\nu}{}^{b}{}_{a} + D_{\mu}
            \tilde{\Lambda}^{b}{}_{a}\ ,
    \end{align}
where $v^{\nu} = -2i \bar{\epsilon}_{2} \GUp{\nu} \epsilon_{1}$ and $\tilde{\Lambda}^{b}{}_{a} = i \bar{\epsilon}_{2} \G{JK} \epsilon_{1} X^{J}_{c} X^{K}_{d} f^{cdb}{}_{a}$. %as found in \cite{Bagger:2007jr}.

In considering the scalars we find a term, $2 i \omega
\bar{\epsilon}_{2} \GUp{MNOPIJ} \epsilon_{1} X^{J}_{a} \tilde{G}_{MNOP}$,
which can be transformed into an object with two gamma matrix
indices by utilizing the self-duality of the flux. We find that the
scalars close into a translation plus a gauge transformation and an
SO(8) R-symmetry,
    \begin{align}
  \nonumber      [ \tilde{\delta}_{1} , \tilde{\delta}_{2} ] X^{I}_{a} &= [ \delta_{1} + \delta_{1}' , \delta_{2} + \delta_{2}' ] X^{I}_{a} \\
        &= v^{\mu} D_{\mu} X^{I}_{a} + \tilde{\Lambda}^{b}{}_{a} X^{I}_{b}  + i R^{I}{}_{J}
        X^{J}_{a}\ ,
    \end{align}
where $R^{I}{}_{J} = 48 \omega \bar{\epsilon}_{2} \GUp{MN}
\epsilon_{1} \tilde{G}_{MNIJ}$ is the $R$-symmetry.

Finally we examine the closure of the Fermions. We find again a term
incorporating $\GUp{(6)}$ which can be converted to $\GUp{(2)}$
using self-duality of $\tilde{G}$. Continuing, we find
    \begin{eqnarray}
        [ \tilde{\delta}_{1} , \tilde{\delta}_{2} ] \Psi_{a} &=& [ \delta_{1} + \delta_{1}' , \delta_{2} + \delta_{2}' ] \Psi_{a} \\
        &=& v^{\mu} D_{\mu} \Psi_{a} + \tilde{\Lambda}^{b}{}_{a} \Psi_{b} + i ( \bar{\epsilon}_{2} \G{\mu} \epsilon_{1} ) \GUp{\mu} E_{\Psi}' - \frac{i}{4} ( \bar{\epsilon}_{2} \G{JK} \epsilon_{1} ) \GUp{JK} E_{\Psi}' \nonumber \\
        && + \frac{i}{4} R_{MN} \GUp{MN} \Psi_{a}\ .
    \end{eqnarray}
Here $E'_{\Psi}$ is the mass deformed Fermionic equation of motion,
    \begin{equation} \label{FermionEOM}
        E_{\Psi}' = \GUp{\nu} D_{\nu} \Psi_{a} + \frac{1}{2} \G{IJ} X^{I}_{c} X^{J}_{d} \Psi_{b} f^{cdb}{}_{a} - \omega \GUp{MNOP} \Psi_{a} \tilde{G}_{MNOP}\ .
    \end{equation}
Consequently, we find that on-shell
    \begin{equation}
        [ \tilde{\delta}_{1} , \tilde{\delta}_{2} ] \Psi_{a} = v^{\mu} D_{\mu} \Psi_{a} + \tilde{\Lambda}^{b}{}_{a} \Psi_{b} + \frac{i}{4} R_{MN} \GUp{MN} \Psi_{a}\ .
    \end{equation}

We also verify that the Fermionic equation of motion maps to the
Bosonic equations of motion under the supersymmetry transformations.
From the proposed mass deformed Lagrangian the scalar equation of
motion is
    \begin{equation}
        E_{X}' = D^{2} X^{I}_{a} - \frac{i}{2} \bar{\Psi}_{c} \GUp{IJ}   X^{J}_{d} \Psi_{b} f^{cdb}{}_{a} - \frac{\partial V}{\partial X^{Ia}} - m^{2} X^{I}_{a} -4c X^{J}_{c} X^{K}_{d} X^{L}_{b} f^{cdb}{}_{a} \tilde{G}_{IJKL} = 0\ .
    \end{equation}
The equation of motion for the gauge field is unchanged and is given
by
    \begin{equation}
        E'_{\tilde{A}} = \tilde{F}_{\mu \nu}{}^{b}{}_{a} + \varepsilon_{\mu \nu \lambda} ( X^{J}_{c} D^{\lambda} X^{J}_{d} + \frac{i}{2} \bar{\Psi}_{c} \GUp{\lambda} \Psi_{d} ) f^{cdb}{}_{a} = 0\ .
    \end{equation}
 Taking the variation of the Fermionic equation of motion \eqref{FermionEOM} gives
    \begin{eqnarray}
        0 &=& \GUp{I} \G{\lambda} X^{I}_{b} E'_{\tilde{A}} \epsilon + \GUp{I} E_{X}' \epsilon \nonumber \\
        && + \frac{96i\omega}{6} \left(-1 + \frac{c}{4\omega} - \star \right) \tilde{G}_{LIJK} \GUp{L} \epsilon X^{I}_{c} X^{J}_{d} X^{K}_{b} f^{cdb}{}_{a} \nonumber \\
        && + \left( m^{2} - \omega^{2} \GUp{MNOP} \GUp{WXYZ} \tilde{G}_{WXYZ} \tilde{G}_{MNOP} \right) \GUp{I} \epsilon X^{I}_{a}\ .
    \end{eqnarray}
Therefore consistency of the equations of motion under supersymmetry
again implies that the conditions \eqref{1stOrderCondition} must be satisfied.

Let us summarize our results. The Lagrangian
    \begin{eqnarray}
        \mathcal{L} &=& - \frac{1}{2} \, {\rm Tr} ( D_{\mu} X^{I} , D^{\mu} X^{I} ) + \frac{i}{2} \, {\rm Tr} ( \bar{\Psi} \GUp{\mu} , D_{\mu} \Psi ) + \frac{i}{4} \, {\rm Tr} ( \bar{\Psi} \G{IJ} , [ X^{I} , X^{J} , \Psi ] ) \nonumber \\
        &&- V - \mathcal{L}_{CS} - \frac{1}{2} m^{2} \delta_{IJ} \, {\rm Tr} ( X^{I} , X^{J} ) - \frac{i c}{16} \, {\rm Tr} ( \bar{\Psi} \GUp{IJKL} , \Psi ) \tilde{G}_{IJKL} \nonumber \\
        &&+ c \, {\rm Tr} ( [ X^{I} , X^{J} , X^{K} ] , X^{L} ) \tilde{G}_{IJKL}
    \end{eqnarray}
is invariant under the supersymmetries
    \begin{align}
        \delta X^{I}_{a} &= i \bar{\epsilon} \GUp{I} \Psi_{a} \nonumber \\
        \delta \tilde{A}_{\mu}{}^{b}{}_{a} &= i \bar{\epsilon} \G{\mu} \G{I} X^{I}_{c} \Psi_{d} f^{cdb}{}_{a}  \\
        \delta \Psi_{a} &= D_{\mu}X^{I}_{a}\GUp{\mu} \GUp{I} \epsilon - \frac{1}{6} X^{I}_{b} X^{J}_{c} X^{K}_{d} f^{bcd}{}_{a} \GUp{IJK} \epsilon + \frac{c}{8}\GUp{IJKL} \GUp{M} \epsilon X^{M}_{a} \tilde{G}_{IJKL} \nonumber
    \end{align}
provided $\tilde{G}_{IJKL}$ is self-dual and satisfies (\ref{Gcon}).
Moreover the supersymmetry algebra closes according to
    \begin{align}
        [ \delta_{1} , \delta_{2} ] \tilde{A}_{\mu}{}^{b}{}_{a} &= v^{\nu} \tilde{F}_{\mu\nu}{}^{b}{}_{a} + D_{\mu} \tilde{\Lambda}^{b}{}_{a}   \nonumber \\
        [ \delta_{1} , \delta_{2} ] X^{I}_{a} &= v^{\mu} D_{\mu} X^{I}_{a} + \tilde{\Lambda}^{b}{}_{a} X^{I}_{b}  + i R^{I}{}_{J} X^{J}_{a} \\
        [ \delta_{1} , \delta_{2} ] \Psi_{a} &= v^{\mu} D_{\mu} \Psi_{a} + \tilde{\Lambda}^{b}{}_{a} \Psi_{b} + \frac{i}{4} R_{MN} \GUp{MN} \Psi_{a}\ . \nonumber
    \end{align}

Taking %
\begin{equation}
G = \mu (dx^3\wedge dx^4\wedge dx^5\wedge dx^6+dx^7\wedge dx^8\wedge dx^9\wedge dx^{10})
\end{equation}
readily leads
to the mass-deformed Lagrangian of \cite{Hosomichi:2008qk}\cite{Gomis:2008cv}.

\section{\sl ${\cal N}=6$ Theories}

Let us now consider the more general case of ${\cal N}=6$
supersymmetry and in particular the ABJM \cite{Aharony:2008ug} and
ABJ \cite{Aharony:2008gk} models which describe an arbitrary number
of M2-branes in an ${\mathbb R}^8/{\mathbb Z}_k$ orbifold. We will use the
notation and conventions of \cite{Bagger:2008se}. Since the
discussion is similar in spirit to the ${\cal N}=8$ case we will
shorten our discussion and largely just present the results of our calculations.

\subsection{\sl Non-Abelian Couplings to Background Fluxes}

In the ${\cal N}=6$ theories there are 4 complex scalars $Z^A$ and
their complex conjugates ${\bar Z}_A$. These are defined in terms of the
spacetime coordinates through
\begin{eqnarray}
  \nonumber  Z^1&=& \frac{1}{\sqrt{2T_{M2}}}(x^3+ix^4)\qquad
  Z^2 =  \frac{1}{\sqrt{2T_{M2}}}(x^5+ix^6)\\
\nonumber   Z^3 &=&  \frac{1}{\sqrt{2T_{M2}}}(x^7-ix^9)\qquad
 Z^4 = \frac{1}{\sqrt{2T_{M2}}}(x^8-ix^{10})\ .
\end{eqnarray}
In particular we will take the formulation in
\cite{Bagger:2008se}.
The scalars and Fermions are endowed with a
triple product $[Z^A,Z^B;{\bar Z}_C]$ or $[{\bar Z}_A,{\bar Z}_b;Z^C]$  and an inner-product
${\rm Tr}{(\bar Z}_A, Z^B)$ subject to a quadratic fundamental identity as well as the
condition
${\rm Tr}({\bar Z}_D, [Z^A,Z^B;{\bar Z}_C])^\star =-{\rm Tr}({\bar Z}_A, [Z^C,Z^D;{\bar Z}_B])$.
To obtain the ABJM/ABJ models  \cite{Aharony:2008gk}\cite{Aharony:2008ug} one should let the fields be $m\times n$ matrices and define
\begin{equation}
[Z^A,Z^B;\bar Z_C] = \lambda(Z^A\bar Z_C^\dag Z^B-Z^B\bar Z_C^\dag
Z^A)\ .
\end{equation}
where $\lambda$ is an arbitrary (but quantized) coupling constant. As such the gauge invariant terms
always involve an equal number of $Z$ and $\bar Z$ coordinates.
Again this is consistent with the interpretation that the M2-branes
are in an ${\mathbb C}^4/{\mathbb Z}_k$ orbifold which acts as $Z^A \to e^{\frac{2\pi i}{k}}
Z^A$.

Following the discussion of the previous section we start with
\begin{eqnarray}
\nonumber   S_{C} &=& \frac{1}{3!}\epsilon^{\mu\nu\lambda}\int d^3 \, x
\Big(
NT_{M2} C_{\mu\nu\lambda}+\frac{3}{2} C_{\mu}{}^A{}_B \, {\rm Tr}
  (D_\nu{\bar Z}_A,D_\lambda Z^B)\\
\nonumber && +\frac{3}{2} C_{\mu}{}_A{}^B \, {\rm Tr}
  (D_\nu Z^A,D_\lambda \bar Z_B)\\
\nonumber && +\frac{3c}{2}C_{\mu\nu AB}{}^{CD} \, {\rm Tr}(
  [D_\lambda {\bar Z}_D,[Z^A,Z^B;{\bar Z}_C])\\
  && +\frac{3c}{2}C_{\mu\nu}{}^{AB}{}_{CD} \, {\rm Tr}(
  [D_\lambda Z^D,[\bar Z_A,\bar Z_B;Z^C])
\Big)\ .
\end{eqnarray}
Integrating by parts we again find a non-gauge invariant term proportional to $\epsilon^{\mu\nu\lambda }\tilde
F_{\nu\lambda}C_{\mu}{}^A{}_B$ which is cancelled by adding
\begin{equation}
S_F = \frac{1}{8}\epsilon^{\mu\nu\lambda} \int d^3 x \, C_{\mu}{}^A{}_B \, {\rm Tr}
  ({\bar Z}_A,\tilde F_{\nu\lambda} Z^B)+C_{\mu}{}_A{}^B \, {\rm Tr}
  ({Z}^A,\tilde F_{\nu\lambda} \bar Z_B)\ .
\end{equation}
As with the case above we also must add
\begin{eqnarray}
S_{CG} &=& -\frac{c}{8\cdot 3!}\epsilon^{\mu\nu\lambda}\int d^3 x \, (C_3\wedge G_4)_{\mu\nu
AB}{}^{CD} \, {\rm Tr}(
  {\bar Z}_D,[Z^A,Z^B;{\bar Z}_C])
\end{eqnarray}
to ensure that the last term is gauge invariant. Thus in total we
have
\begin{eqnarray}
\nonumber   S_{flux} &=& S_C+S_F+S_{CG}\\
\nonumber &=&
\frac{1}{3!}\epsilon^{\mu\nu\lambda}\int d^3 x \,
\Big(
NT_{M2} C_{\mu\nu\lambda}\\&&+\frac{3}{4} G_{\mu\nu}{}^A{}_B \, {\rm Tr}
  ({\bar Z}_A,D_\lambda Z^B)+\frac{3}{4} G_{\mu\nu}{}_A{}^B \, {\rm Tr}
  ({Z}^A,D_\lambda \bar Z_B)\\
\nonumber  && -\frac{c}{4}(G_7+\frac{1}{2}C_3\wedge G_4)_{\mu\nu\lambda AB}{}^{CD}\, {\rm Tr}(
  [{\bar Z}_D,[Z^A,Z^B;{\bar Z}_C])
\Big) \, .
\end{eqnarray}

\subsection{\sl Supersymmetry}

Following on as before we wish to supersymmetrize the action
    \begin{equation}
        \mathcal{L} = \mathcal{L}_{{\cal N}=6} + \mathcal{L}_{mass} + \mathcal{L}_{flux}\ ,
    \end{equation}
where ${\cal L}_{{\cal N}=6}$ is the ${\cal N}=6$
Chern-Simons-Matter
Lagrangian.
We restrict to backgrounds where
\begin{equation}\label{Sflux2}
\mathcal{L}_{flux} = \frac{c}{4} \, {\rm Tr}( [\bar Z_D,[ Z^A, Z^B;\bar Z_C]) \tilde G_{AB}{}^{CD} \ ,
\end{equation}
with
\begin{eqnarray}
\nonumber \tilde G_{AB}{}^{CD} &=&-
\frac{1}{3!}\epsilon^{\mu\nu\lambda}(G_7+\frac{1}{2}C_3\wedge G_{4})_{\mu\nu\lambda
AB}{}^{CD}\\ &=&
\frac{1}{4}\epsilon_{ABEF}\epsilon^{CDGH}G^{EF}{}_{GH}\ .
\end{eqnarray}
Finally we take the ansatz for $\mathcal{L}_{mass}$ to be
\begin{equation}
\mathcal{L}_{mass} = - m^2 \, {\rm Tr}( \bar Z_A , Z^A ) + b \, {\rm Tr}(\bar \psi^A , \psi_F ) \tilde G_{AE}{}^{EF}\ .
\end{equation}
We propose the following modification to the Fermion supersymmetry variation
\begin{equation}
\delta ' \psi_{Ad} = \omega \epsilon_{DF} Z^F_d \tilde G_{AE}{}^{ED}\ ,
\end{equation}
where $\omega$ is a real parameter.

After applying the supersymmetry transformations to $\mathcal{L}$ we find that taking $b=-i \omega$ eliminates the covariant derivative terms. The terms that are second order in $\tilde{G}$ must vanish separately and this gives the condition
\begin{equation}\label{Gcon2}
\tilde{G}_{AE}{}^{EB} \tilde{G}_{BF}{}^{FC} = \frac{m^2}{\omega^2} \delta_A^C\ .
\end{equation}

The remaining terms in the variation are
\begin{eqnarray}
\nonumber \delta \mathcal{L} &=& + 2 i \omega \, {\rm Tr} ( \bar Z_D , [ \bar \psi_F \epsilon^{DA} , Z^Q ; \bar Z_Q ] ) \tilde{G}_{AE}{}^{EF}\\
\nonumber   && + i \omega \, {\rm Tr} ( \bar Z_D , [ \bar \psi_F \epsilon^{QD} , Z^A ; \bar Z_Q ] ) \tilde{G}_{AE}{}^{EF} \\
    && + 2 i \omega \, {\rm Tr}( \bar Z_D , [ \bar \psi_K \epsilon^{AD}, Z^K ; \bar Z_F ] ) \tilde{G}_{AE}{}^{EF}\\
\nonumber   && + \frac{ic}{2} \, {\rm Tr}( \bar Z_D , [ \bar \psi_K \epsilon^{AK}, Z^B ; \bar Z_C ] ) \tilde{G}_{AB}{}^{CD}\\
\nonumber   && + \frac{i \omega}{2} \varepsilon^{AKQD} \varepsilon_{IJFP} \, {\rm Tr}( \bar Z_D , [ \bar \psi_K \epsilon^{IJ},Z^P ; \bar Z_Q ] ) \tilde{G}_{AE}{}^{EF}\\
\nonumber   && + \text{c.c.}\ ,
\end{eqnarray}
where we have made use of the reality condition $\epsilon_{FP} = \frac{1}{2} \varepsilon_{IJFP} \,
\epsilon^{IJ}$. To proceed we need to restrict $\tilde{G}$ to have the form
\begin{equation}\label{Gid}
\tilde{G}_{AB}{}^{CD} = \frac{1}{2} \delta^C_B \tilde{G}_{AE}{}^{ED} - \frac{1}{2} \delta^C_A
\tilde{G}_{BE}{}^{ED} - \frac{1}{2} \delta^D_B \tilde{G}_{AE}{}^{EC} + \frac{1}{2} \delta^D_A
\tilde{G}_{BE}{}^{EC}\ ,
\end{equation}
with $\tilde{G}_{AE}{}^{EA}=0$. Substituting for $\tilde{G}_{AB}{}^{CD}$ allows us to factor out the common
term
 ${\rm Tr} ( \bar Z_D , [ \bar \psi_K \epsilon^{IJ} , Z^P ; \bar Z_Q ] ) \tilde{G}_{AE}{}^{EF}$. This factor is
separately anti-symmetric in $IJ$ and $DQ$ so after expanding out $\varepsilon^{AQKD} \varepsilon_{IJFP} = 4!
\delta^{[ AQKD ]}_{\phantom{[} IJFP}$ we have
\begin{eqnarray}
\nonumber \delta \mathcal{L} &=& i \omega \left( \frac{c}{2\omega} -2 \right) ( \delta^A_I \delta^K_J \delta ^D_F \delta^Q_P + \delta^K_I \delta^Q_J \delta ^D_F \delta^A_P )\\
&& \times \, {\rm Tr} ( \bar Z_D , [ \bar \psi_K \epsilon^{IJ}, Z^P ; \bar Z_Q ] ) \tilde{G}_{AE}{}^{EF}\\
\nonumber && + \text{c.c.}
\end{eqnarray}
Therefore the Lagrangian is invariant under supersymmetry if $\omega=c/4$. Taking the trace of equation (\ref{Gcon2})
allows us to deduce that
\begin{equation}
m^2 = \frac{1}{32\cdot 4!}c^2 G^2
\end{equation}
where $G^2 = 6G_{AB}{}^{CD} G^{AB}{}_{CD}=12 G_{AE}{}^{EB}G_{BF}{}^{FA}$.

In examining the closure of the superalgebra we find
\begin{eqnarray}
\left[ \delta_1,\delta_2 \right] \tilde{A}_{\mu}{}^c{}_d &=& v^{\nu} \tilde{F}_{\mu\nu}{}^c{}_d +D_{\mu} (\Lambda_{\bar a b} f^{cb \bar a}{}_d)\\
\left[ \delta_1,\delta_2 \right] Z^A_d &=& v^{\mu} D_{\mu} Z^A_d + \Lambda_{\bar c b} f^{ab \bar c}{}_d Z^A_a -
i R^A{}_B Z^B_d - i Y Z^A_d
\end{eqnarray}
where
\begin{eqnarray}
v^{\mu} &=& \frac{i}{2} \bar \epsilon_2^{CD} \gamma^{\mu} \epsilon^1_{CD}\\
\Lambda_{\bar c b} &=& i ( \bar \epsilon_2^{DE} \epsilon^1_{CE} - \bar \epsilon_1^{DE} \epsilon^2_{CE} ) \bar Z_{D \bar c} Z^C_b\\
R^A{}_B &=& \omega \left( ( \bar \epsilon_1^{AC} \epsilon^2_{DB} - \bar \epsilon_2^{AC} \epsilon^1_{DB} ) - \frac{1}{4} ( \bar \epsilon_1^{EC} \epsilon^2_{DE} - \bar \epsilon_2^{EC} \epsilon^1_{DE} ) \delta^A_B \right) \tilde{G}_{CM}{}^{MD}\\
Y &=& \frac{\omega}{4} ( \bar \epsilon_1^{EC} \epsilon^2_{DE} - \bar \epsilon_2^{EC} \epsilon^1_{DE} ) \tilde{G}_{CM}{}^{MD}\ .
\end{eqnarray}
Acting with the commutator on the Fermions gives
\begin{eqnarray}
\nonumber [ \delta_1 , \delta_2 ] \psi_{Dd} &=& v^\mu D_\mu \psi_{Dd} +
\Lambda_{\bar a b}f^{cb \bar a}{}_d\psi_{Dc}\\
\nonumber &&-\frac{i}{2}(\bar\epsilon_1^{AC}\epsilon_{2AD}-\bar\epsilon_2^{AC}\epsilon_{1AD})E'_{Cd}\\
\nonumber && +\frac{i}{4}(\bar\epsilon^{AB}_1\gamma_\nu\epsilon_{2AB})\gamma^\nu
E'_{Dd}\\
&& + i R^A{}_D \psi_{Ad} - i Y \psi_{Dd}
\end{eqnarray}
provided the 4-form satisfies $\tilde{G}_{AE}{}^{EA}=0$. The new Fermionic equation of motion is
\begin{eqnarray}
\nonumber E'_{Cd} &=& \gamma^{\mu} D_{\mu} \psi_{Cd} +f^{ab \bar c}{}_d \psi_{Ca}
Z^D_b \bar Z_{D \bar c} -2 f^{ab \bar c}{}_d \psi_{Da} Z^D_b \bar Z_{C \bar c} \\
&& - \varepsilon_{CDEF} f^{ab \bar c}{}_d\psi^D_{\bar c} Z^E_a Z^F_b + \frac{c}{4} \tilde{G}_{CE}{}^{EB}
\psi_{Bd}\ .
\end{eqnarray}
Consistency of the Bosonic and Fermionic equations of motion under supersymmetry requires that $\tilde{G}_{AE}{}^{EB} \tilde{G}_{BF}{}^{FC} = \frac{m^2}{\omega^2} \delta^C_A$, which is the same condition as found in demonstrating invariance of the action.

Choosing $\tilde G_{AB}{}^{CD}$ to have the form (\ref{Gid}) with
\begin{equation}\label{ex2}
\tilde G_{AB}{}^{BC} = \left(
                  \begin{array}{cccc}
                    \mu & 0 &0&0\\
                    0 & \mu &0&0\\
                    0 & 0 &-\mu&0\\
                     0& 0 &0&-\mu\\
                  \end{array}
                \right)\ ,
\end{equation}
gives the mass-deformed Lagrangian of \cite{Hosomichi:2008jb}\cite{Gomis:2008vc}.

\section{\sl Background Curvature}

Our final point is to understand the physical origin of the mass-squared term in the effective action which is
quadratic in the masses. Note that this term is a simple, $SO(8)$-invariant mass term for all the scalar
fields. Furthermore it does not depend on any non-Abelian features of the theory. Therefore we can derive this
term by simply considering a single M2-brane and compute the unknown constant $c$.

We can understand the origin of this term as follows. We have seen
that it arises as a consequence of supersymmetry. For a single
M2-brane supersymmetry arises as a consequence of $\kappa$-symmetry
and $\kappa$-symmetry is valid whenever an M2-brane is propagating
in a background that satisfies the equations of motion of
eleven-dimensional supergravity \cite{Bergshoeff:1987cm}.

The multiple M2-brane actions implicitly assume that the background
is simply flat space or an orbifold thereof. However the inclusion
of a non-trivial flux  implies that there is now a source for the
eleven-dimensional metric which is of order flux-squared. Thus for
there to be $\kappa$-supersymmetry and hence supersymmetry it
follows that the background must be curved. This in turn will lead
to a potential in the effective action of an M2-brane. In particular
given a 4-form flux $G_4$ the Bosonic equations of
eleven-dimensional supergravity are
\begin{eqnarray}
 \nonumber R_{mn} - \frac{1}{2}g_{mn}R &=& \frac{1}{2\cdot 3!}G_{mpqr}G_n{}^{pqr} - \frac{1}{4\cdot 4!}g_{mn}G^2 \\
   d\star G_4 -\frac{1}{2}G_4\wedge G_4 &=&0\ .
\end{eqnarray}
At lowest order in fluxes we see that $g_{mn}=\eta_{mn}$ and $G_4$ is constant. However at second order there
are source terms. To start with we will assume that, at lowest order, only $G_{IJKL}$ is non-vanishing. To
solve these equations we introduce a non-trivial metric of the form
\begin{equation}
g_{mn} = \left(
           \begin{array}{cc}
             e^{2\omega}\eta_{\mu\nu} & 0 \\
             0 & g_{IJ} \\
           \end{array}
         \right)\ ,
\end{equation}
where $\omega = \omega(x^I) = \omega( X^I/T_{M2}^{\frac{1}{2}})$ and
$g_{IJ} = g_{IJ}(x^I) = g_{IJ}( X^I/T_{M2}^{\frac{1}{2}})$.

Let us look at an M2-brane in this background. The first term in the action (\ref{BSTaction}) is
\begin{eqnarray}
\nonumber  S_1 &=& -T_{M2}\int d^3x  \sqrt{-{\rm det}(e^{2\omega}\eta_{\mu\nu}+\partial_\mu x^I\partial_\nu x^J g_{IJ})} \\
   &=&-T_{M2}\int d^3x \, e^{3\omega}\left(1 + \frac{1}{2}e^{-2\omega}\partial_\mu x^I\partial^\mu
   x^Jg_{IJ}+\ldots\right)\\
   \nonumber &=& -\int d^3x \left(T_{M2}e^{3\omega} + \frac{1}{2}e^{\omega}\partial_\mu X^I\partial^\mu
   X^Jg_{IJ}+\ldots\right) \ .
\end{eqnarray}
Next we note that, in the decoupling limit $T_{M2}\to \infty$, we
can expand
\begin{equation}
e^{2\omega(x)} = e^{2\omega(X^I/\sqrt{T_{M2}})} =
1+\frac{2}{T_{M2}}\omega_{IJ}X^IX^J+\ldots\ ,
\end{equation}
and
\begin{equation}
g_{IJ}(x) = g_{IJ}( X^I/\sqrt{T_{M2}}) = \delta_{IJ}+\ldots\ ,
\end{equation}
so that
\begin{equation}
S_1 =-\int d^3x \left(T_{M2} + 3\omega_{IJ}X^IX^J + \frac{1}{2}\partial_\mu X^I\partial^\mu
   X^J\delta_{IJ}+\ldots\right)\ ,
\end{equation}
where the ellipsis denotes terms that vanish as $T_{M2}\to \infty$. Thus we see that in the decoupling limit we
obtain the mass term for the scalars. Similar mass terms for M2-branes were also studied in
\cite{Skenderis:2003da}  for pp-waves.

To compute the warp-factor $\omega$ we can expand $g_{mn} =
\eta_{mn} + h_{mn}$, where $h_{mn}$ is second order in the fluxes,
and linearize the Einstein equation. If we impose the gauge
$\partial^m h_{mn} -\frac{1}{2}\partial_n h^p{}_p=0$ then Einstein's
equation becomes
\begin{equation}
-\frac{1}{2}\partial_p\partial^p \left(h_{mn} -\frac{1}{2}\eta_{mn}
h_q{}^q \right)= \frac{1}{2\cdot 3!}G_{mpqr}G_n{}^{pqr} -
\frac{1}{4\cdot 4!}g_{mn}G^2\ .
\end{equation}
This reduces to two coupled sets of equations corresponding to
choosing  indices $(m,n)=(\mu,\nu)$ and $(m,n)=(I,J)$. Contracting
the latter with $\delta^{IJ}$ one finds that $h_I{}^I = 4h_p{}^p$
and hence $h_p{}^p = -\frac{1}{3}h_\mu{}^\mu$. With this in hand the
$(m,n)=(\mu,\nu)$ terms in Einstein's equation reduce to
\begin{equation}
\partial_I\partial^I e^{2\omega} = \frac{1}{3\cdot 4!} G^2\ ,
\end{equation}
and hence, to leading order in the fluxes,
\begin{equation}
e^{2\omega} = 1+\frac{1}{48\cdot4!}G^2 \delta_{IJ}x^Ix^J\ ,
\end{equation}
so that $S_1$ contributes the term
\begin{equation}
S_1 = -\int d^3 x \, \frac{1}{32\cdot4!}G^2X^2
\end{equation}
to the potential.

Next we must look at the second, Wess-Zumino term, in (\ref{BSTaction});
\begin{equation}
S_2 = \frac{T_{M2}}{3!}\int d^3 x \, \epsilon^{\mu\nu\lambda} C_{\mu\nu\lambda}\ .
\end{equation}
Although we have assumed that $C_{\mu\nu\lambda}=0$ at leading order, the $C$-field equation of motion implies
that $G_{I \mu\nu\lambda }=\partial_ I C_{\mu\nu\lambda}$  is second order in $G_{IJKL}$. In particular if we
write $C_{\mu\nu\lambda} = C_0\epsilon_{\mu\nu\lambda}$ we find, assuming $G_{IJKL}$ is self-dual,  the
equation
\begin{equation}
\partial_I\partial^I C_0 = \frac{1}{2\cdot 4!}G^2\ ,
\end{equation}
where $G^2 = G_{IJKL}G^{IJKL}$. The solution is
\begin{equation}
C_0=\frac{1}{32\cdot 4!}G^2 \delta_{IJ}x^Ix^J\ .
\end{equation}
Thus we find that $S_2$ gives a second contribution to the scalar potential
\begin{equation}
S_2 = -\int d^3 x \, \frac{1}{32\cdot4!}G^2X^2\ .
\end{equation}
Note that this is equal to the scalar potential derived from $ S_1$. Therefore if we were to break
supersymmetry and consider anti-M2-branes, where the sign of the Wess-Zumino term changes, we would not find a
mass for the scalars.

In total we find the mass-squared
\begin{equation}
m^2= \frac{1}{8\cdot 4!}G^2 \ .
\end{equation}
Comparing with (\ref{Gcon}) we see that $c^2=4$, {\it e.g.} $c=2$. Note that we have performed this
calculation using the notation of the ${\cal N}=8$ theory, however a similar calculation also holds in the
${\cal N}=6$ case with the same result.

\section{\sl Conclusions}

In this paper we discussed the coupling of multiple M2-branes with ${\cal N}=6,8$ supersymmetry to the
background gauge fields of eleven-dimensional supergravity. In particular we gave
a local and gauge invariant form for the `Myers terms' in the limit $M_{pl}\to \infty$. We supersymmetrized
these flux terms in the case
where the fluxes preserve the supersymmetry and Lorentz symmetry of
M2-branes to obtain the  massive models of \cite{Hosomichi:2008qk}\cite{Gomis:2008cv}\cite{Hosomichi:2008jb}\cite{Gomis:2008vc}. We also showed how the
flux-squared term in the effective action, which arises as a
mass term for the scalar fields, is generated through a back
reaction of the fluxes on the eleven-dimensional geometry.

The results we have found using gauge invariance fit naturally with the ${\mathbb R}^8/{\mathbb Z}_k$ orbifold
interpretation of the background. However for the ${\cal N}=6$ theories with $k=1,2$ the orbifold action is
less restrictive and this allows for additional terms. In particular for $k=2$ we expect terms where total
number of $Z^A$ and $\bar Z_B$ fields are even (but not necessarily equal). In addition for $k=1$ there should
be terms with any number of $Z^A$ and $\bar Z_B$ fields. Such terms are not gauge invariant on their own but
presumably can be made so by including monopole operators which, for $k=1,2$, are local.

\section*{\sl Acknowledgements}

We would like to thank Greg Moore,  Rob Myers and Costis Papageorgakis for insightful discussions. This work is
supported in part by the PPARC rolling grant PP/C5071745/1 the EU Marie Curie Research Training Grant
MRTN-CT-2004-512194 and STFC rolling grant ST/G000/395/1.

\end{document}